\documentclass[aps,preprint,amsmath,amssymb]{revtex4-1}

\usepackage{dcolumn}
\usepackage{bm}
\usepackage{hyperref}

\begin{document}

\title{Extension of Malcev algebra and applications to gravity}

\author{Junpei Harada}
\email[]{jharada@hoku-iryo-u.ac.jp}
\affiliation{Research Center for Higher Education, Health Sciences University of Hokkaido, Japan}
\date{June 4, 2019}

\begin{abstract}
We investigate extensions of Malcev algebras and give an explicit example of extended algebras.
We present a new algebraic identity, which can be regarded as a generalization of the Jacobi identity or the Malcev identity.
As applications to gravity, we demonstrate that the extended algebra can be linked with general relativity.
\end{abstract}

\maketitle

\section{Introduction\label{sec:intro}}
Nonassociative algebras play a significant role in physics. 
An introduction to nonassociative physics is given in~\cite{Okubo:1990nv}, and recent developments are presented in~\cite{Szabo:2019hhg}.
An elementary example of a nonassociative algebra is a Lie algebra.
However, a Lie algebra is too restrictive in some systems. 
For instance, it is known that other nonassociative algebras appear in quantum mechanics with magnetic monopoles, string theory, and M-theory~\cite{Szabo:2019hhg,Gunaydin:2013nqa}.

A Malcev algebra~\cite{Malcev:1955}(or Moufang-Lie algebra by Malcev himself) is an important example of nonassociative algebras other than Lie algebras.
A Malcev algebra is a generalization of a Lie algebra in the sense that any Lie algebra is a Malcev algebra but the converse is not true. 
A Malcev algebra plays an important role~\cite{Gunaydin:2013nqa}, however, there still remains possible other nonassociative algebras beyond Malcev algebras may play some essential role. 

In the present paper we investigate extensions of Malcev algebras and present an explicit example of extended algebras.
We propose a new algebraic identity, which can be regarded as a generalization of the Jacobi identity for Lie algebras, or the Malcev identity for Malcev algebras.
Because a new identity is a generalization of the Jacobi identity or the Malcev identity, we can expect that it has more chances to be relevant to broader physical systems.
Then, as applications to gravity, we demonstrate that the extended algebra can be linked with general relativity. 

The paper is organized as follows.
In section~\ref{sec:eMalcev}, we investigate extensions of Malcev algebras.
In particular, a new algebraic identity is given.
In section~\ref{sec:app}, we present applications of the extended algebras to gravity. 
Section~\ref{sec:sum} is devoted to a summary.

\section{Extension of Malcev algebra\label{sec:eMalcev}}
\subsection{Deformation of the Poincar\'e algebra\label{subsec:Poincare}}

We begin with the Poincar\'e algebra in $3+1$ dimensions. 
We denote the Lorentz generators as $J_{ab}$, and the translations as $P_a$, with $a, b = 0, 1, 2, 3$.
The Poincar\'e algebra is defined by 
\begin{subequations}
\begin{align}
	\left[J_{ab}, J_{cd}\right] &= - \eta_{ac}J_{bd} + \eta_{bc}J_{ad} + \eta_{ad}J_{bc} - \eta_{bd}J_{ac}, \label{eq:JJ}\\
	\left[J_{ab},P_c\right] &= - \eta_{ac} P_b + \eta_{bc} P_a,\label{eq:JP}\\
	\left[P_a, P_b\right] &=0, \label{eq:PP}
\end{align}
\end{subequations}
where $\eta_{ab}$ is a metric with a signature $(-, +, +, +)$, and $[\cdot, \cdot]$ is a Lie bracket.
The Poincar\'e algebra is a Lie algebra, that is, eqs.~\eqref{eq:JJ}-\eqref{eq:PP} satisfy the following Jacobi identity,
\begin{align}
	[[x,y],z] + [[y,z],x] + [[z,x],y] =0,
	\label{eq:Jacobi-id}
\end{align}
where $x$, $y$, and $z$ are any elements in the algebra.

In this paper we study an extension of Lie algebras to non-Lie algebras in which the Jacobi identity is violated. 
As a minimal example, we set the deformation of eq.~\eqref{eq:PP},
\begin{align}
	\left[P_a, P_b\right]&=\frac{\lambda}{2} \epsilon_{abcd} J^{cd}.
	\label{eq:PP2}
\end{align}
Here, $\epsilon_{abcd}$ is a totally anti-symmetric Levi-Civita symbol with $\epsilon_{0123}=1$, and $\lambda$ is a physical constant with dimensions of $(\mbox{Length})^{-2}$. 
If the value of $\lambda$ is sufficiently small, it is difficult to experimentally distinguish between eqs.~\eqref{eq:PP2} and~\eqref{eq:PP}. 
However, as long as $\lambda$ is non-vanishing, eq.~\eqref{eq:PP2} yields a non-Lie algebra. 
We will find that the algebra thus obtained is a generalization of Malcev algebras. 

We find that the following identities hold,
\begin{subequations}
\begin{align}
	[[J_{ab}, J_{cd}], J_{ef}]+[[J_{cd},J_{ef}],J_{ab}]+ [[J_{ef}, J_{ab}],J_{cd}]&=0,\label{eq:JJJ1}\\
	[[J_{ab}, J_{cd}], P_e]+ [[J_{cd}, P_e], J_{ab}] + [[P_e, J_{ab}], J_{cd}]&=0, \label{eq:JJP1}\\
	[[J_{ab}, P_c], P_d]+ [[P_c, P_d], J_{ab}] + [[P_d, J_{ab}], P_c]&= 0,\label{eq:JPP1}\\
	[[P_a, P_b], P_c]+ [[P_b, P_c],P_a] +[[P_c, P_a],P_b]&= -3\lambda\epsilon_{abcd}P^d. \label{eq:PPP1}
\end{align}
\end{subequations}
These identities are explained as follows.
Eqs.~\eqref{eq:JJJ1} and~\eqref{eq:JJP1} are irrelevant to eq.~\eqref{eq:PP2}, and therefore the Jacobi identities are valid. 
In contrast, eqs.~\eqref{eq:JPP1} and \eqref{eq:PPP1} are relevant to eq.~\eqref{eq:PP2}, and therefore they are rather nontrivial.
A calculation shows that although each term in eq.~\eqref{eq:JPP1} does not vanish (it vanishes in the case of the Poincar\'e algebra), the Jacobi identity is valid.

Eq.~\eqref{eq:PPP1} is exceptional and interesting. 
Eqs.~\eqref{eq:JP} and~\eqref{eq:PP2} imply that each term in the left-hand side of eq.~\eqref{eq:PPP1} is identical as follows,
\begin{align}
	[[P_a, P_b], P_c] = [[P_b, P_c],P_a] = [[P_c, P_a],P_b] = -\lambda\epsilon_{abcd}P^d. \label{eq:PPPeq} 
\end{align}
Furthermore, eq.~\eqref{eq:PPPeq} and an anti-commutativity of a bracket $[\cdot, \cdot]$ imply that the following identity holds,
\begin{align}
	[[P_a, P_b], P_c] = -[P_a, [P_b, P_c]]. \label{eq:anti-associative} 
\end{align}
In abstract algebra, eq.~\eqref{eq:anti-associative} is called the \emph{anti-associative law}. 
This anti-associativity leads to a violation of the Jacobi identity.

\subsection{Malcev algebras\label{subsec:Malcev}}
We have confirmed that eqs.~\eqref{eq:JJ}, \eqref{eq:JP}, and \eqref{eq:PP2} do not satisfy the Jacobi identity, that is, the deformed algebra is not a Lie algebra. 
Then, it is natural to ask whether it is a Malcev algebra, which is a generalization of Lie algebras.  

A Malcev algebra is defined as an algebra that satisfies the following Malcev identity~\cite{Malcev:1955},
\begin{align}
		[[x,y], [x,z]]
		= [[[x,y],z],x] + [[[y,z],x],x] + [[[z,x],x],y].
		\label{eq:Malcev-id1}
\end{align}
Eq.~\eqref{eq:Malcev-id1} is a generalization of the Jacobi identity as follows.
Here, it is convenient to define a ternary bracket $[\cdot, \cdot, \cdot]$ by 
\begin{align}
	[x, y, z] := [[x, y],z] + [[y, z],x] + [[z, x],y], \label{eq:db-bra}
\end{align}
which is totally anti-symmetric in $x$, $y$, and $z$. 
We can rewrite eq.~\eqref{eq:Malcev-id1} in terms of a ternary bracket $[\cdot, \cdot, \cdot]$ as~\cite{Gunaydin:2013nqa} 
	\begin{align}
		[x, y, [z,x]] + [[x, y, z],x] = 0.
		\label{eq:Malcev-id2}
	\end{align}
\emph{Proof of eq.~\eqref{eq:Malcev-id2}};
 \begin{align}
	0&= -[[x,y],[x,z]] + [[[x,y],z],x] + [[[y,z],x],x] + [[[z,x],x],y]  \nonumber\\		 
	&=[[x,y],[z,x]] + [[[x,y],z],x] + [[[y,z],x],x] + [[[z,x],x],y] + [[y,[z,x]],x] - [[y,[z,x]],x]  \nonumber\\
	&=[[x,y],[z,x]] + [[y,[z,x]],x] + [[[z,x],x],y] + [[[x,y],z] + [[y,z],x] + [[z,x],y],x]\nonumber\\
	&=[x,y,[z,x]] + [[x,y,z],x].
\end{align}
	\begin{flushright}$\Box$\end{flushright}
If an algebra is a Lie algebra, that is $[x, y, z]=0$, then it implies the validity of eq.~\eqref{eq:Malcev-id2}.
Therefore, any Lie algebra is a Malcev algebra, and in this sense the Malcev identity is a generalization of the Jacobi identity.

However, we find that eqs.~\eqref{eq:JJ}, \eqref{eq:JP}, and \eqref{eq:PP2} do not satisfy the Malcev identity. 
For instance, we set $x=P_0$, $y=P_1$, and $z=P_2$ in eq.~\eqref{eq:Malcev-id2}, and then we have
\begin{align}
	[P_0, P_1, [P_2,P_0]] + [[P_0, P_1, P_2],P_0]&= -\lambda[P_0, P_1, J_{31}] -3\lambda[P_3, P_0] \nonumber\\
	&= 0 -3\lambda(-\lambda J_{12})= 3\lambda^2 J_{12} \not= 0.
	\label{eq:violation_Malcev}
\end{align}
Thus, we have learned that eqs.~\eqref{eq:JJ}, \eqref{eq:JP}, and \eqref{eq:PP2} do not satisfy either the Jacobi identity or the Malcev identity.
In other words, the deformed algebra is not either a Lie algebra or a Malcev algebra. 
In section~\ref{subsec:extension_Malcev}, we present a new identity that the deformed algebra obeys.

\subsection{Extension of Malcev algebras\label{subsec:extension_Malcev}}
In this paper we present the following new identity, 
\begin{align}
	[[x,y],[[z,x],y]]=[[[[x,y],z],x],y] + [[[[y,z],x],x],y] + [[[[z,x],y],y],x].\label{eq:gMalcev-id1}	
\end{align}
It should be noted that each term includes {\it four} brackets (the Jacobi identity has {\it two} brackets, and the Malcev identity includes {\it three} brackets). 
We show that eq.~\eqref{eq:gMalcev-id1} is a generalization of the Malcev identity, and then we will find that the deformed algebra satisfies eq.~\eqref{eq:gMalcev-id1}. 

First, we show that any Lie algebra satisfies eq.~\eqref{eq:gMalcev-id1}. 
This can be proved as follows. 
We can rewrite eq.~\eqref{eq:gMalcev-id1} in terms of a ternary bracket $[\cdot, \cdot, \cdot]$ defined by eq.~\eqref{eq:db-bra} as,
 \begin{align}
	[x,y,[[x,z],y]]+[[[x,y,z],x],y]&=0. &\label{eq:gMalcev-id2}
 \end{align} 
\emph{Proof of eq.~\eqref{eq:gMalcev-id2}};
 \begin{align}
 	0&=-[[x,y],[[z,x],y]] + [[[[x,y],z],x],y] + [[[[y,z],x],x],y] + [[[[z,x],y],y],x] \nonumber\\
	 &=[[x,y],[[x,z],y]] + [[[[x,y],z],x],y] + [[[[y,z],x],x],y] + [[y,[[x,z],y]],x]\nonumber\\
	 &=[[x,y],[[x,z],y]]+ [[y,[[x,z],y]],x] + [[[[x,y],z],x],y] + [[[[y,z],x],x],y] \nonumber\\
	 &\quad + [[[[z,x],y],x],y] - [[[[z,x],y],x],y] \nonumber\\
	 &=[[x,y],[[x,z],y]] + [[y,[[x,z],y]],x] + [[[[x,z],y],x],y] + [[[x,y,z],x],y]  \nonumber\\
	 &=[x,y,[[x,z],y]] + [[[x,y,z],x],y].
 \end{align}
 \begin{flushright}$\Box$\end{flushright}
If an algebra is a Lie algebra, that is $[x, y, z]=0$, then it implies the validity of eq.~\eqref{eq:gMalcev-id2}. 
Therefore, any Lie algebra satisfies eq.~\eqref{eq:gMalcev-id2}.

Furthermore, we show that any Malcev algebra satisfies eq.~\eqref{eq:gMalcev-id1}.
This can be proved as follows.
For any Malcev algebra, we can rewrite the second term in eq.~\eqref{eq:gMalcev-id2},
\begin{align}
	&[[x,y,z],x],y] = -[[x, y, [z,x]],y] = [[y, x, [z,x]],y] = - [y, x, [[z,x],y]] = - [x,y,[[x,z],y]], \label{eq:second}
\end{align}
where we have used eq.~\eqref{eq:Malcev-id2}. 
Eq.~\eqref{eq:second} implies the validity of eq.~\eqref{eq:gMalcev-id2}. 
Therefore, any Malcev algebra satisfies eq.~\eqref{eq:gMalcev-id2}.
Thus, the new identity, eq.~\eqref{eq:gMalcev-id1}, is a generalization of the Malcev identity. 

Here, we find that the following identities hold,
\begin{subequations}
 \begin{align}
	[J_{ab},J_{cd},[[J_{ab},J_{ef}],J_{cd}]]+[[[J_{ab},J_{cd},J_{ef}],J_{ab}],J_{cd}]&=0, \label{eq:newJJJ}\\
	[J_{ab},J_{cd},[[J_{ab},P_e],J_{cd}]]+[[[J_{ab},J_{cd},P_e],J_{ab}],J_{cd}]&=0, \label{eq:newJJP}\\
	[J_{ab},P_c,[[J_{ab},J_{de}],P_c]]+[[[J_{ab},P_c,J_{de}],J_{ab}],P_c]&=0, \label{eq:newJPJ}\\
	[J_{ab},P_c,[[J_{ab},P_d],P_c]]+[[[J_{ab},P_c,P_d],J_{ab}],P_c]&=0, \label{eq:newJPP}\\
	[P_a,J_{bc},[[P_a,J_{de}],J_{bc}]]+[[[P_a,J_{bc},J_{de}],P_a],J_{bc}]&=0, \label{eq:newPJJ}\\
	[P_a,J_{bc},[[P_a,P_d],J_{bc}]]+[[[P_a,J_{bc},P_d],P_a],J_{bc}]&=0, \label{eq:newPJP}\\
	[P_a,P_b,[[P_a,J_{cd}],P_b]]+[[[P_a,P_b,J_{cd}],P_a],P_b]&=0, \label{eq:newPPJ}\\	
	[P_a,P_b,[[P_a,P_c],P_b]]+[[[P_a,P_b,P_c],P_a],P_b]&=0.	\label{eq:newPPP}
\end{align}
\end{subequations}
These identities are explained as follows. 
The validity of eqs.~\eqref{eq:newJJJ}-\eqref{eq:newPPJ} is trivial, since eqs.~\eqref{eq:JJJ1}-\eqref{eq:JPP1} imply that $[J_{ab},x,y]=0$ for all $x$ and $y$. 
Only eq.~\eqref{eq:newPPP} is nontrivial. 
To demonstrate the validity of eq.~\eqref{eq:newPPP}, we set $x=P_0$, $y=P_1$, and $z=P_2$ in eq.~\eqref{eq:newPPP} (it is straightforward to calculate other $x$, $y$, and $z$), and then we find that 
\begin{align}
	[P_0,P_1,[[P_0,P_2],P_1]]+[[[P_0,P_1,P_2],P_0],P_1]
	&=\lambda[P_0,P_1,[J_{31},P_1]] - 3\lambda[[P_3,P_0],P_1]\nonumber\\
	&=\lambda[P_0,P_1,P_3]  +3\lambda^2 [J_{12},P_1] \nonumber\\
	&=3\lambda^2 P_2 + 3\lambda^2 (-P_2) 
	=0.
\end{align}
Thus, eqs.~\eqref{eq:JJ}, \eqref{eq:JP}, and \eqref{eq:PP2} satisfy eq.~\eqref{eq:gMalcev-id2}.
The identity, eq.~\eqref{eq:gMalcev-id1}, defines an extension of Malcev algebras, just as the Jacobi identity defines Lie algebras, or just as the Malcev identity defines Malcev algebras. 

Before going to the next section, we give a comment on a relation to a Lie 3-algebra (or 3-Lie algebra). 
We can rewrite eqs.~\eqref{eq:JJJ1}-\eqref{eq:PPP1} in terms of a ternary bracket $[\cdot, \cdot, \cdot]$ defined by eq.~\eqref{eq:db-bra},
\begin{subequations}
\begin{align}
	[J_{ab}, J_{cd}, J_{ef}] &=0,\label{eq:JJJ2}\\
	[J_{ab}, J_{cd}, P_e] &=0, \label{eq:JJP2}\\
	[J_{ab}, P_c, P_d] &= 0, \label{eq:JPP2}\\
	[P_a, P_b, P_c] &= -3\lambda\epsilon_{abcd}P^d. \label{eq:PPP2}
\end{align}
\end{subequations}
Interestingly, eq.~\eqref{eq:PPP2} has appeared in the literature~\cite{Okubo:1992qt,Kawamura:2002yz,Kawamura:2003cw,Bagger:2006sk,Gustavsson:2007vu,Bagger:2007jr,Bagger:2007vi,Ho:2008bn,Ho:2016hob} as a ternary system without regarding eq.~\eqref{eq:db-bra}. 
A ternary system equipped with eq.~\eqref{eq:PPP2} is called a Lie 3-algebra ${\cal A}_4$~\cite{Ho:2008bn} or a quaternionic triple system~\cite{Okubo:1992qt}. 
In those literature, a ternary system equipped with eq.~\eqref{eq:PPP2} has been studied as a ternary system, in which a binary operation denoted by a bracket $[\cdot, \cdot]$ is not necessarily defined. 
In contrast, in the present paper, eq.~\eqref{eq:PPP2} is a consequence of eqs.~\eqref{eq:JJ}, \eqref{eq:JP}, \eqref{eq:PP2}, and \eqref{eq:db-bra}, which define binary operations on the algebra. 

\section{Applications to gravity\label{sec:app}}
In this section we present applications of the extended algebra to gravity.
We begin with a theory of the spin connection and the vierbein in four dimensions.
We will denote the Lorentz indices as $a, b, c, d$, the spacetime indices as $\mu, \nu, \rho, \sigma$, and the Minkowski metric as $\eta_{ab}$, of signature ($-$ $+$ $+$ $+$). 

The spin connection ${\omega_\mu}^{ab}$ and the vierbein ${e_\mu}^a$ can be regarded as components of a single gauge field $A_\mu$, 
\begin{align}
	A_\mu = \frac{1}{2} {\omega_\mu}^{ab} J_{ab} + {e_\mu}^a P_a.
	\label{gauge}
\end{align}
We then calculate the curvature tensor
\begin{align}
	F_{\mu\nu} 
	&= \partial_\mu A_\nu - \partial_\nu A_\mu + \left[A_\mu, A_\nu \right] \nonumber\\
	&= \frac{1}{2}\left(\partial_\mu {\omega_\nu}^{ab} - \partial_\nu {\omega_\mu}^{ab}+ {\omega_{\mu}}^{ac} {\omega_{\nu c}}^{b} - {\omega_\nu}^{ac} {\omega_{\mu c}}^{b} \right)J_{ab} \nonumber \\
	& \quad +  \left(\partial_\mu {e_\nu}^a - \partial_\nu {e_\mu}^a+ {\omega_\mu}^{ab} e_{\nu b} - {\omega_\nu}^{ab} e_{\mu b} \right) P_a \nonumber\\
	& \quad + \frac{1}{2} \left( {e_\mu}^a {e_\nu}^b - {e_\nu}^a {e_\mu}^b\right) [P_a, P_b].
	\label{curvature1}
\end{align}
Here, the coefficient of $J_{ab}$ is the curvature tensor for the Lorentz group,
\begin{eqnarray}
	{R_{\mu\nu}}^{ab} 
	= \partial_\mu {\omega_\nu}^{ab} - \partial_\nu {\omega_\mu}^{ab} 
	+ {\omega_\mu}^{ac} {\omega_{\nu c}}^b - {\omega_\nu}^{ac} {\omega_{\mu c}}^b,
	\label{Lcurvature}
\end{eqnarray}
and the coefficient of $P_a$ is the torsion tensor,
\begin{eqnarray}
	{T_{\mu\nu}}^a 
	= \partial_\mu {e_\nu}^a - \partial_\nu {e_\mu}^a 
	+ {\omega_\mu}^{ab} e_{\nu b} - {\omega_\nu}^{ab} e_{\mu b}. 
	\label{Torsion}
\end{eqnarray}
Substituting eq.~(\ref{eq:PP2}) into eq.~(\ref{curvature1}), we have
\begin{eqnarray}
	F_{\mu\nu} 
	= \frac{1}{2} {R_{\mu\nu}}^{ab} J_{ab}
	+ {T_{\mu\nu}}^a P_a
	+ \frac{\lambda}{2} \epsilon_{abcd} {E_{\mu\nu}}^{ab}J^{cd},
	\label{curvature2}
\end{eqnarray}
where
\begin{align}
	{E_{\mu\nu}}^{ab} = \frac{1}{2}\left( {e_\mu}^a {e_\nu}^b - {e_\nu}^a {e_\mu}^b\right).
\end{align}

For an infinitesimal gauge parameter $u$, the gauge transformation of the gauge field should be 
\begin{eqnarray}
	\delta A_\mu 
	= \left[A_\mu, u\right] + \partial_\mu u.
	\label{deltaA1}
\end{eqnarray}
Here, a gauge parameter $u$ is a zero-form $u=\frac{1}{2} \tau^{ab} J_{ab} + \rho^a P_a$, with $\tau$ and $\rho$ being the infinitesimal parameters. 
Furthermore, since the Jacobi identity is violated, the transformation law of $[A_\mu, A_\nu]$ should be
\begin{align}
	\delta ([A_\mu, A_\nu]) = [\delta A_\mu, A_\nu] + [A_\mu, \delta A_\nu] + [A_\mu, A_\nu, u],
\end{align}
which implies that $\delta$ is not a derivation. Then, we have
\begin{eqnarray}
	\delta F_{\mu\nu} 
	= \left[ F_{\mu\nu}, u \right].
	\label{dF2}
\end{eqnarray}
Here, we set $\tau^{ab}=0$ for simplicity, since a Lorentz invariance is trivial. 
Then, we calculate
\begin{align}
	 \delta F_{\mu\nu}
	&= \left( {R_{\mu\nu}}^{ab} 
	+ \frac{\lambda}{2} \epsilon^{abcd} \left( e_{\mu c} e_{\nu d}  - e_{\nu c} e_{\mu d} \right) \right) \rho_b P_a +\frac{\lambda}{2} \epsilon_{abcd} {T_{\mu\nu}}^a \rho^b J^{cd}.
	\label{dF3}
\end{align}

In order to construct the Einstein-Hilbert action, we set the following symmetric bilinear form,
\begin{subequations}
\begin{align}
	&\left< J_{ab}, J_{cd} \right> = \eta_{ac}\eta_{bd} - \eta_{bc}\eta_{ad},\\
	&\left< J_{ab}, P_{c} \right> =0,\\
	&\left< P_{a}, P_{b} \right> = 0.
	\label{BF2}
\end{align}
\end{subequations}
This is not invariant under the gauge transformation, however, it will be shown that the variation of the action $\delta S$ is proportional to ${T_{\mu\nu}}$ which vanishes.

Using this form, we obtain the action $S=\int \langle F, F\rangle$,
\begin{align}
	S &= \frac{1}{8} \int \epsilon^{\mu\nu\rho\sigma} \left( {R_{\mu\nu}}^{ab} R_{\rho\sigma ab} \right) d^4 x
	 + \frac{\lambda}{4} \int \epsilon^{\mu\nu\rho\sigma} \epsilon_{abcd} \left( {e_\mu}^a {e_\nu}^b {R_{\rho\sigma}}^{cd} \right) d^4 x
\end{align}
or simply as
\begin{eqnarray}
	S = & 
	\frac{1}{2} \int \left( R^{ab} \wedge R_{ab} \right)
	 + \frac{\lambda}{2} \int \epsilon_{abcd} \left( e^a \wedge e^b \wedge R^{cd} \right). 
	 \label{gravity}
\end{eqnarray}
The first term in eq.~(\ref{gravity}) is the Pontryagin topological invariant, and the second term is the Einstein-Hilbert term. 
The field equations are obtained from this action; variation with respect to the vierbein leads to the Einstein equation in vacuum, and variation with respect to the spin connection leads to the field equation, ${T_{\mu\nu}}^a=0$. 

Here, the field equation ${T_{\mu\nu}}^a=0$ is algebraic in ${\omega_\mu}^{ab}$, which implies that ${\omega_\mu}^{ab}$ can be described in terms of ${e_\mu}^a$. 
Substituting ${\omega_\mu}^{ab}$ written in terms of ${e_\mu}^a$ into the action, we can go to  the second order formalism that contains only the vierbein ${e_\mu}^a$. 
We calculate the variation of the action under the transformation generated by $P^a$,
\begin{align}
	\delta S &= \frac{\lambda}{4} \int \epsilon^{\mu\nu\rho\sigma} \epsilon_{abcd}
	 {T_{\mu\nu}}^a \rho^b \left( {R_{\rho\sigma}}^{cd} 	+ \frac{\lambda}{2}  \epsilon^{cdef} \left( e_{\rho e} e_{\sigma f} - e_{\sigma e} e_{\rho f} \right) \right) d^4 x.
	\label{deltaS}
\end{align}
Thus, we find that $\delta S$ is proportional to ${T_{\mu\nu}}^a$ which vanishes.

\section{Summary\label{sec:sum}}
In this paper we have investigated some algebraic properties of extension of Malcev algebra. 
In particular, we have proposed a new algebraic identity given by eq.~\eqref{eq:gMalcev-id1}.
We have shown that eq.~\eqref{eq:gMalcev-id1} can be regarded as a generalization of the Jacobi identity for Lie algebras, or a generalization of the Malcev identity for Malcev algebras. 
Then, we have presented applications of the algebra to gravity. 
The algebra studied in this paper has a potential for other applications, just as Lie algebras or Malcev algebras.

\bibliography{references}

\begin{thebibliography}{13}%
\makeatletter
\providecommand \@ifxundefined [1]{%
 \@ifx{#1\undefined}
}%
\providecommand \@ifnum [1]{%
 \ifnum #1\expandafter \@firstoftwo
 \else \expandafter \@secondoftwo
 \fi
}%
\providecommand \@ifx [1]{%
 \ifx #1\expandafter \@firstoftwo
 \else \expandafter \@secondoftwo
 \fi
}%
\providecommand \natexlab [1]{#1}%
\providecommand \enquote  [1]{``#1''}%
\providecommand \bibnamefont  [1]{#1}%
\providecommand \bibfnamefont [1]{#1}%
\providecommand \citenamefont [1]{#1}%
\providecommand \href@noop [0]{\@secondoftwo}%
\providecommand \href [0]{\begingroup \@sanitize@url \@href}%
\providecommand \@href[1]{\@@startlink{#1}\@@href}%
\providecommand \@@href[1]{\endgroup#1\@@endlink}%
\providecommand \@sanitize@url [0]{\catcode `\\12\catcode `\$12\catcode
  `\&12\catcode `\#12\catcode `\^12\catcode `\_12\catcode `\%12\relax}%
\providecommand \@@startlink[1]{}%
\providecommand \@@endlink[0]{}%
\providecommand \url  [0]{\begingroup\@sanitize@url \@url }%
\providecommand \@url [1]{\endgroup\@href {#1}{\urlprefix }}%
\providecommand \urlprefix  [0]{URL }%
\providecommand \Eprint [0]{\href }%
\providecommand \doibase [0]{http://dx.doi.org/}%
\providecommand \selectlanguage [0]{\@gobble}%
\providecommand \bibinfo  [0]{\@secondoftwo}%
\providecommand \bibfield  [0]{\@secondoftwo}%
\providecommand \translation [1]{[#1]}%
\providecommand \BibitemOpen [0]{}%
\providecommand \bibitemStop [0]{}%
\providecommand \bibitemNoStop [0]{.\EOS\space}%
\providecommand \EOS [0]{\spacefactor3000\relax}%
\providecommand \BibitemShut  [1]{\csname bibitem#1\endcsname}%
\let\auto@bib@innerbib\@empty
\bibitem [{\citenamefont {Okubo}(2011)}]{Okubo:1990nv}%
  \BibitemOpen
  \bibfield  {author} {\bibinfo {author} {\bibfnamefont {S.}~\bibnamefont
  {Okubo}},\ }\href
  {http://www.cambridge.org/mw/academic/subjects/physics/theoretical-physics-and-mathematical-physics/introduction-octonion-and-other-non-associative-algebras-physics?format=AR}
  {\emph {\bibinfo {title} {{Introduction to Octonion and other Non-Associative
  Algebras in Physics}}}},\ Montroll Memorial Lecture Series in Mathematical
  Physics\ (\bibinfo  {publisher} {Cambridge Univ. Press},\ \bibinfo {address}
  {Cambridge, UK},\ \bibinfo {year} {2011})\BibitemShut {NoStop}%
\bibitem [{\citenamefont {Szabo}(2019)}]{Szabo:2019hhg}%
  \BibitemOpen
  \bibfield  {author} {\bibinfo {author} {\bibfnamefont {R.~J.}\ \bibnamefont
  {Szabo}},\ }in\ \href@noop {} {\emph {\bibinfo {booktitle} {{18th Hellenic
  School and Workshops on Elementary Particle Physics and Gravity (CORFU2018)
  Corfu, Corfu, Greece, August 31-September 28, 2018}}}}\ (\bibinfo {year}
  {2019})\ \Eprint {http://arxiv.org/abs/1903.05673} {arXiv:1903.05673
  [hep-th]} \BibitemShut {NoStop}%
\bibitem [{\citenamefont {Gnaydin}\ and\ \citenamefont
  {Minic}(2013)}]{Gunaydin:2013nqa}%
  \BibitemOpen
  \bibfield  {author} {\bibinfo {author} {\bibfnamefont {M.}~\bibnamefont
  {Gnaydin}}\ and\ \bibinfo {author} {\bibfnamefont {D.}~\bibnamefont
  {Minic}},\ }\href {\doibase 10.1002/prop.201300010} {\bibfield  {journal}
  {\bibinfo  {journal} {Fortsch. Phys.}\ }\textbf {\bibinfo {volume} {61}},\
  \bibinfo {pages} {873} (\bibinfo {year} {2013})},\ \Eprint
  {http://arxiv.org/abs/1304.0410} {arXiv:1304.0410 [hep-th]} \BibitemShut
  {NoStop}%
\bibitem [{\citenamefont {Malcev}(1955)}]{Malcev:1955}%
  \BibitemOpen
  \bibfield  {author} {\bibinfo {author} {\bibfnamefont {A.~I.}\ \bibnamefont
  {Malcev}},\ }\href@noop {} {\bibfield  {journal} {\bibinfo  {journal} {Math.
  Sb.}\ }\textbf {\bibinfo {volume} {36(78)}},\ \bibinfo {pages} {569}
  (\bibinfo {year} {1955})}\BibitemShut {NoStop}%
\bibitem [{\citenamefont {Okubo}(1993)}]{Okubo:1992qt}%
  \BibitemOpen
  \bibfield  {author} {\bibinfo {author} {\bibfnamefont {S.}~\bibnamefont
  {Okubo}},\ }\href {\doibase 10.1063/1.530076} {\bibfield  {journal} {\bibinfo
   {journal} {J. Math. Phys.}\ }\textbf {\bibinfo {volume} {34}},\ \bibinfo
  {pages} {3273} (\bibinfo {year} {1993})},\ \Eprint
  {http://arxiv.org/abs/hep-th/9212051} {arXiv:hep-th/9212051 [hep-th]}
  \BibitemShut {NoStop}%
\bibitem [{\citenamefont {Kawamura}(2003{\natexlab{a}})}]{Kawamura:2002yz}%
  \BibitemOpen
  \bibfield  {author} {\bibinfo {author} {\bibfnamefont {Y.}~\bibnamefont
  {Kawamura}},\ }\href {\doibase 10.1143/PTP.109.153} {\bibfield  {journal}
  {\bibinfo  {journal} {Prog. Theor. Phys.}\ }\textbf {\bibinfo {volume}
  {109}},\ \bibinfo {pages} {153} (\bibinfo {year} {2003}{\natexlab{a}})},\
  \Eprint {http://arxiv.org/abs/hep-th/0207054} {arXiv:hep-th/0207054 [hep-th]}
  \BibitemShut {NoStop}%
\bibitem [{\citenamefont {Kawamura}(2003{\natexlab{b}})}]{Kawamura:2003cw}%
  \BibitemOpen
  \bibfield  {author} {\bibinfo {author} {\bibfnamefont {Y.}~\bibnamefont
  {Kawamura}},\ }\href {\doibase 10.1143/PTP.110.579} {\bibfield  {journal}
  {\bibinfo  {journal} {Prog. Theor. Phys.}\ }\textbf {\bibinfo {volume}
  {110}},\ \bibinfo {pages} {579} (\bibinfo {year} {2003}{\natexlab{b}})},\
  \Eprint {http://arxiv.org/abs/hep-th/0304149} {arXiv:hep-th/0304149 [hep-th]}
  \BibitemShut {NoStop}%
\bibitem [{\citenamefont {Bagger}\ and\ \citenamefont
  {Lambert}(2007)}]{Bagger:2006sk}%
  \BibitemOpen
  \bibfield  {author} {\bibinfo {author} {\bibfnamefont {J.}~\bibnamefont
  {Bagger}}\ and\ \bibinfo {author} {\bibfnamefont {N.}~\bibnamefont
  {Lambert}},\ }\href {\doibase 10.1103/PhysRevD.75.045020} {\bibfield
  {journal} {\bibinfo  {journal} {Phys. Rev.}\ }\textbf {\bibinfo {volume}
  {D75}},\ \bibinfo {pages} {045020} (\bibinfo {year} {2007})},\ \Eprint
  {http://arxiv.org/abs/hep-th/0611108} {arXiv:hep-th/0611108 [hep-th]}
  \BibitemShut {NoStop}%
\bibitem [{\citenamefont {Gustavsson}(2009)}]{Gustavsson:2007vu}%
  \BibitemOpen
  \bibfield  {author} {\bibinfo {author} {\bibfnamefont {A.}~\bibnamefont
  {Gustavsson}},\ }\href {\doibase 10.1016/j.nuclphysb.2008.11.014} {\bibfield
  {journal} {\bibinfo  {journal} {Nucl. Phys.}\ }\textbf {\bibinfo {volume}
  {B811}},\ \bibinfo {pages} {66} (\bibinfo {year} {2009})},\ \Eprint
  {http://arxiv.org/abs/0709.1260} {arXiv:0709.1260 [hep-th]} \BibitemShut
  {NoStop}%
\bibitem [{\citenamefont {Bagger}\ and\ \citenamefont
  {Lambert}(2008{\natexlab{a}})}]{Bagger:2007jr}%
  \BibitemOpen
  \bibfield  {author} {\bibinfo {author} {\bibfnamefont {J.}~\bibnamefont
  {Bagger}}\ and\ \bibinfo {author} {\bibfnamefont {N.}~\bibnamefont
  {Lambert}},\ }\href {\doibase 10.1103/PhysRevD.77.065008} {\bibfield
  {journal} {\bibinfo  {journal} {Phys. Rev.}\ }\textbf {\bibinfo {volume}
  {D77}},\ \bibinfo {pages} {065008} (\bibinfo {year} {2008}{\natexlab{a}})},\
  \Eprint {http://arxiv.org/abs/0711.0955} {arXiv:0711.0955 [hep-th]}
  \BibitemShut {NoStop}%
\bibitem [{\citenamefont {Bagger}\ and\ \citenamefont
  {Lambert}(2008{\natexlab{b}})}]{Bagger:2007vi}%
  \BibitemOpen
  \bibfield  {author} {\bibinfo {author} {\bibfnamefont {J.}~\bibnamefont
  {Bagger}}\ and\ \bibinfo {author} {\bibfnamefont {N.}~\bibnamefont
  {Lambert}},\ }\href {\doibase 10.1088/1126-6708/2008/02/105} {\bibfield
  {journal} {\bibinfo  {journal} {JHEP}\ }\textbf {\bibinfo {volume} {02}},\
  \bibinfo {pages} {105} (\bibinfo {year} {2008}{\natexlab{b}})},\ \Eprint
  {http://arxiv.org/abs/0712.3738} {arXiv:0712.3738 [hep-th]} \BibitemShut
  {NoStop}%
\bibitem [{\citenamefont {Ho}\ \emph {et~al.}(2008)\citenamefont {Ho},
  \citenamefont {Hou},\ and\ \citenamefont {Matsuo}}]{Ho:2008bn}%
  \BibitemOpen
  \bibfield  {author} {\bibinfo {author} {\bibfnamefont {P.-M.}\ \bibnamefont
  {Ho}}, \bibinfo {author} {\bibfnamefont {R.-C.}\ \bibnamefont {Hou}}, \ and\
  \bibinfo {author} {\bibfnamefont {Y.}~\bibnamefont {Matsuo}},\ }\href
  {\doibase 10.1088/1126-6708/2008/06/020} {\bibfield  {journal} {\bibinfo
  {journal} {JHEP}\ }\textbf {\bibinfo {volume} {06}},\ \bibinfo {pages} {020}
  (\bibinfo {year} {2008})},\ \Eprint {http://arxiv.org/abs/0804.2110}
  {arXiv:0804.2110 [hep-th]} \BibitemShut {NoStop}%
\bibitem [{\citenamefont {Ho}\ and\ \citenamefont {Matsuo}(2016)}]{Ho:2016hob}%
  \BibitemOpen
  \bibfield  {author} {\bibinfo {author} {\bibfnamefont {P.-M.}\ \bibnamefont
  {Ho}}\ and\ \bibinfo {author} {\bibfnamefont {Y.}~\bibnamefont {Matsuo}},\
  }\bibfield  {booktitle} {\emph {\bibinfo {booktitle} {{Proceedings, Nambu
  Memorial Symposium: Osaka, Japan, September 29, 2015}}},\ }\href {\doibase
  10.1093/ptep/ptw075} {\bibfield  {journal} {\bibinfo  {journal} {PTEP}\
  }\textbf {\bibinfo {volume} {2016}},\ \bibinfo {pages} {06A104} (\bibinfo
  {year} {2016})},\ \Eprint {http://arxiv.org/abs/1603.09534} {arXiv:1603.09534
  [hep-th]} \BibitemShut {NoStop}%
\end{thebibliography}%

\end{document}